\documentclass[a4paper,11pt]{article}
\usepackage{pos}

\usepackage{array, booktabs}
\usepackage[demo, export]{adjustbox}  % it also load graphicx
\usepackage[skip=1ex]{caption}

\title{Crowd-sourced particle physics stories from DESY-CMS}
\author[a,b]{Freya Blekman}
\author[a]{Andrea Cardini}
\author*[a]{Lucia Ximena Coll Saravia}

\affiliation[a]{Deutsches Elektronen-Synchrotron DESY, Notkestr. 85, 22607 Hamburg, Germany}

\affiliation[b]{Universit\"at Hamburg, Luruper Chaussee 149, 22761 Hamburg, Germany}

\emailAdd{lucia.ximena.coll.saravia@cern.ch}

\abstract{The CMS at DESY outreach Instagram account (@cmsatdesy) serves as a platform for science communication and outreach for a large experimental particle physics group. The initiative aims to promote scientific research, engage young scientists in outreach activities, and showcase their contributions. Instagram was chosen for its strong alignment with the target demographic and its broad user base in Germany and internationally.

The account highlights the work of young scientists, providing insights into their scientific journeys and disseminating particle physics outreach content. Multiple contributors collaborate on content creation, offering early career researchers opportunities for training in science communication while maintaining a manageable time commitment. This paper presents the evolution of the project, its initial objectives, target audience, and the experiences gained in content development and public engagement on social media platforms.}

\FullConference{%
  42nd International Conference on High Energy physics - ICHEP2024\\
  17-24 July, 2024\\
  Prague, Czech Republic
}

\begin{document}
\maketitle

% READ ME:
% Please, take into consideration that this project is still in PROGRESS 
% and it will be a FIRST DRAFT. 
% A grammar/ortography check will be performed when the draft is completed.

\section{Introduction}
The Compact Muon Solenoid (CMS) \cite{cmsexp} is a versatile detector at the Large Hadron Collider (LHC) at CERN, designed to enhance our understanding of the Standard Model (SM) and explore physics beyond it. The CMS collaboration consists of approximately 4,000 scientists from around 240 institutions worldwide.
The Deutsches Elektronen-Synchrotron (DESY), a leading German research center, boasts a scientific community of about 3,200 members, including 2,700 staff and around 500 PhD students and postdoctoral researchers. The DESY-based members of the CMS Collaboration engage in all major aspects of particle physics research, including detector hardware development and instrumentation, algorithmic development, physics analysis, and data taking.

The CMS team at DESY is a very diverse and large physics group that comprises 77 active scientific members (excluding master's and bachelor's students and interns). The diversity as far as nationality, career status and gender statistics, are presented in Figure \ref{fig:us-stats}.
\iffalse
\begin{figure}[hbt]

\begin{center}
 \begin{tabular}{cc}
\includegraphics[width=0.5\textwidth]{graphs/us-countries.png} & \includegraphics[width=0.4\textwidth]{graphs/us-career.png}\\ 
 & \\
\end{tabular}
\includegraphics[scale=0.35]{graphs/us-gender.png}
\end{center}
 \caption[CMS at DESY statistics.]{Nationality, status of the career and gender statistics of the members of the CMS at DESY team. }
 \label{fig:us-stats}
\end{figure}
\fi
\begin{figure}[hbt]

\begin{center}
 \includegraphics[width=0.35\textwidth]{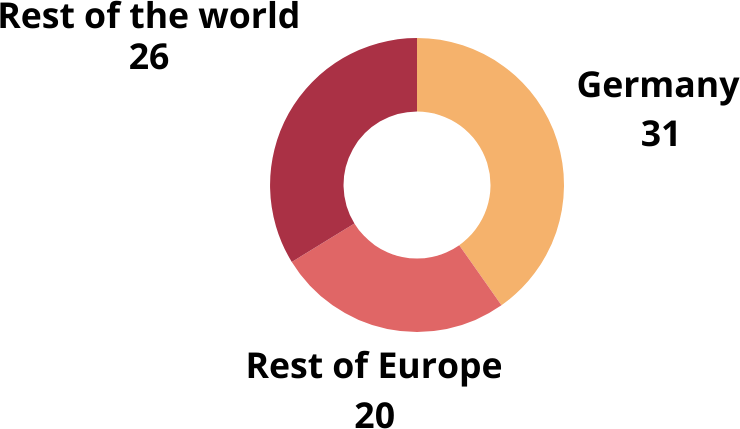}
 \includegraphics[width=0.29\textwidth]{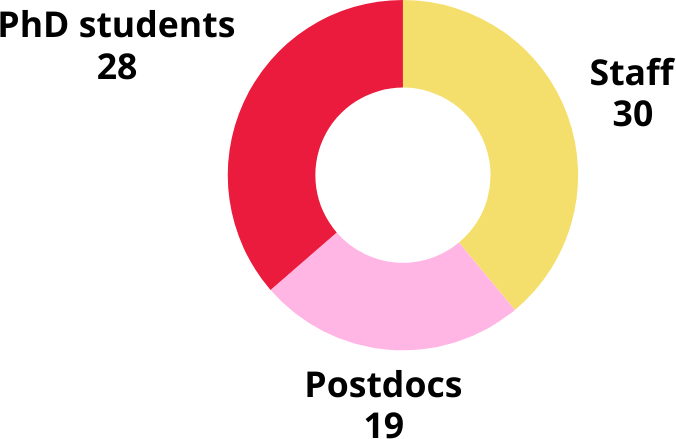} 
\raisebox{0.4cm}{\includegraphics[width=0.25\textwidth]{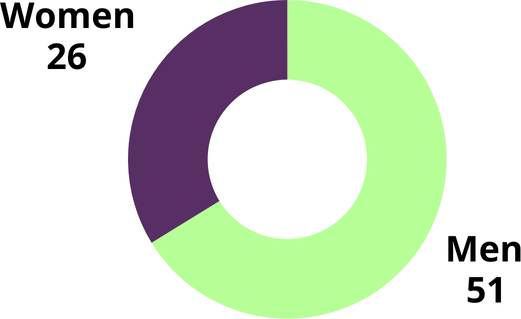}}
\end{center}
 \caption[CMS at DESY statistics.]{Nationality, status of the career and gender statistics of the members of the CMS at DESY team. }
 \label{fig:us-stats}
\end{figure}

As is expected from a national laboratory, DESY and members of the CMS group are actively involved in various outreach activities aimed at introducing particle physics to a broad set of stake holders in the general public. These include diverse outreach activities:
\begin{itemize}
  \setlength\itemsep{0pt}
  \setlength\parsep{0pt}
  \setlength\topsep{0pt}
    \item Talks in bars, typically reaching science-interested audiences aged 30–50+. \cite{wissen}
    \item A diverse program of presentations in schools, engaging students from 4 to 18 years old~\cite{wirwollenswissen}.
    \item Appearances on television programs, which generally attract a very wide and diverse demographic, that is dominated by ages 40 and up~\cite{newspaperstats}.  %\cite{cmsigtv}
    \item Contributing to popular science articles for newspapers, which due to the nature of the medium target the aged 50-70+ audience~\cite{newspaperstats}.
\end{itemize}

These outreach efforts are at a typical level of diversity for most larger institutes in the field. A common gap in most typical outreach efforts is identifiable when reviewing the target audiences: the young adult to mid-30s age group is not reached by these common outreach activities.

\section{Crowd-sourced particle physics stories}

The CMS @ DESY Instagram ~\cite{cmsig} project was specifically designed to reach the 18-34 age group. Instagram was chosen because it is the social media platform where the target audience is most active. In Germany alone, there were over 30 million users in 2023, with more than 50\% of them between 18 and 34 years old~\cite{germanig}. This demographic aligns perfectly with age groups not reached by other outreach activities. Using Instagram as the communication platform targets several goals, as discussed below.

\subsection{Outreach}

Our target audience is likely unfamiliar with the details of particle physics and possibly with physics in general. The goal is to spark genuine interest in the field, capture the attention of the general public, foster curiosity, and create a space for authentic engagement with our research.

To achieve this, short videos that explain basic concepts are produced. Examples of topics are popular science topics that will gain some recognition, such as the LHC, cosmic rays, or the photoelectric effect, using simple language and focusing on fundamental ideas.

\subsection{Science communication}

Excluding Nobel prize winning research, particle physics findings are often not well known outside the internal community. One of the main objectives is to share new results with the general public, demonstrating their significance and utility of day-to-day research, also focusing on its incremental and sometimes more challenging aspects. This factor is important to raise awareness about the more mundane results and developments in our field that occur regularly, not only focusing on once-in-a-lifetime discoveries like the Higgs boson.

There is a wealth of proof that the social media activity of scientists increases the general public's trust in science as a whole~\cite{scientistwhoselfie}. This effect is powerful when scientists are also shown to have interests beyond science and take part in everyday activities, which is why the CMSatDESY account also focuses on social aspects of lab life and anecdotes and hobbies of senior staff members.

This personal approach not only makes the communication more accessible to non-expert audiences but in addition gives public recognition and credit to particle physics for their individual contributions, something that is often impossible in laboratory and collaboration science communication. Institutional accounts from CERN, DESY, the CMS collaboration, and so forth focus on press-related content, and they do not highlight individual groups or members. Making a diverse group of scientists of all ages part of this initiative makes it personal and showcases the diversity of science while highlighting small contributions.

\subsection{Training of local people}

Science communication is one of the underrated skills a scientist will need throughout their career, and it is typically not included at any stage of formal physics education. Mastering this skill can open doors for scientists at all career levels; the ability to communicate their work effectively can lead to more opportunities to stand out in the competitive academic job market, promote new results versus competing scientific efforts, secure grants, and, on a larger scale, convince key stakeholders such as funding agencies and politicians to fund new or ongoing experiments.

The CMSatDESY project provides a safe space for scientists at all career stages to practice and refine their communication skills. The low-risk environment of Instagram encourages exploring creative ways to share science through low-risk elements. The goal is not for team members to all become professional science communicators but rather to equip the majority with the skills necessary to navigate a scientific career while being able to communicate their work effectively. Due to the nature of modern academia, it is expected that most early career researchers will eventually end up working outside universities and national labs. In professional environments, the skills to efficiently communicate complex topics are equally desirable.

\subsection{Recruitment}

As noted before, the target audience is in a crucial age group for recruitment. The science communication leg of CMSatDESY aims to reach not only the science-interested general public but also young people at key decision-making stages in their careers. Specifically, we target final-year bachelor's or master's students who are considering short internships or searching for thesis opportunities, recent graduates looking for PhD positions, and newly graduated PhDs seeking postdoctoral positions. To engage this audience, the short videos that showcase the 'life at the lab' of different members of our group are useful. These videos highlight a typical day in a large institute, featuring activities such as working on a hardware experiment, programming, having lunch, enjoying coffee with the team, attending meetings, and other daily tasks.

\subsection{Organization}
The team working on this project comprises one staff member, one postdoc, and one PhD student 
%(Figure \ref{fig:team})
, the three of them working full time in scientific activities. Their responsibilities include:
\begin{enumerate}
\setlength\itemsep{0pt}
  \setlength\parsep{0pt}
  \setlength\topsep{0pt}
    \item Searching for volunteers with recent results or publications, or willing to contribute with content related to 'life at the lab.'
    \item Collaborating with them to produce the best possible content.
    \item Maintaining a schedule of publications and events for the group.
    \item Managing the account (e.g. posting, answering messages, comments, etc).
\end{enumerate}

The project benefits from crowd-sourcing, which helps distribute the workload more evenly among the whole CMS group and makes it possible for the three team members to work on this project while not neglecting their own scientific responsibilities. This approach also contributes to the project's long-term sustainability, ensuring that it remains viable and effective over time.

%%
%%\begin{figure}[h]
%%\centering
%%\includegraphics[width=0.6\textwidth,trim={0 0 0 3cm},clip]{images/IMG_4622.jpg}
%%\caption[The team]{The team, composed of one staff member, one postdoc, and one PhD student.}
%%\label{fig:team}
%%\end{figure}

\section{Performance}

\begin{figure}[bt]
\includegraphics[width=0.45\textwidth]{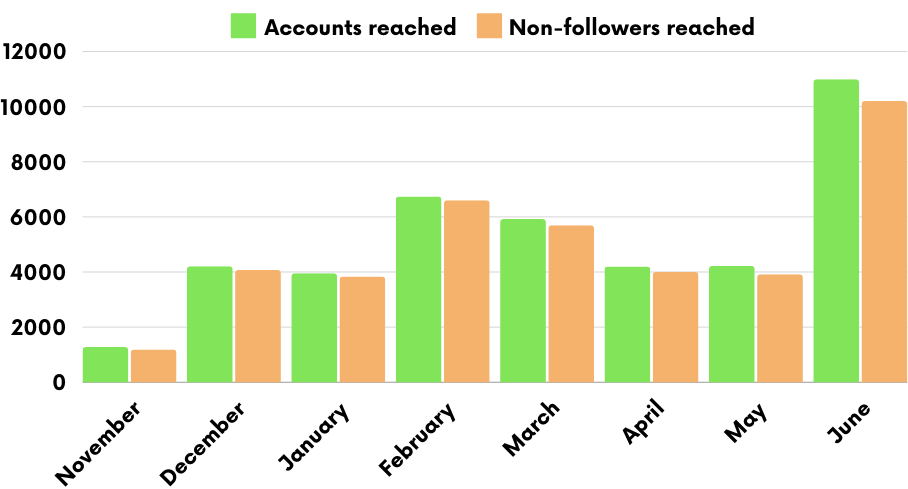}
\includegraphics[width=0.45\textwidth]{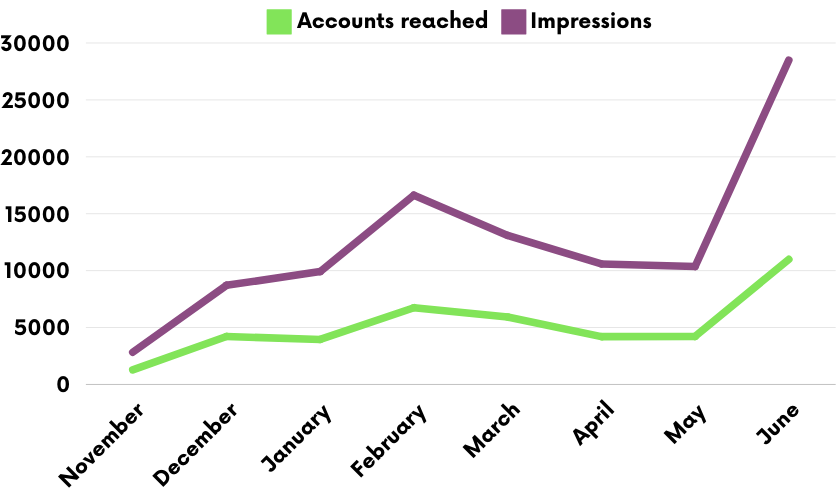} 
\centering
\caption[Accounts reached]{(Left): Comparison between the total accounts reached and the non followers reached from November 2023 to June 2024. (Right):  Comparison between the total accounts reached and the impressions from November 2023 to June 2024.  }
\label{fig:accountsreached}
\end{figure}

This Instagram account was launched in November 2023, which is 10 months prior to the writing of this document. Since then, the account has shown a consistent growth trend. As of now, the majority of the account's reach still comprises non-followers, as characteristic of new accounts. This is illustrated in Figure \ref{fig:accountsreached}. In the figure, the total number of accounts reached is shown in green, while the non-followers reached are depicted in orange. The number of followers reached can be derived as the difference between the total accounts reached and non-followers reached. 

Notably, growth has been significant. We can see with time a general increase in accounts reached, the difference between total and non-followers is widening, we are starting to get more followers engaged. Fluctuations in growth can be attributed to the randomess of social media, as well as the quantity, quality, and type of videos published each month, as well as interactions with similar accounts. For instance, a video posted in collaboration with the CMS experiment account tends to increase visibility. Such posts reach the larger number of CMS experiment account followers and are consequently recommended by Instagram to users interested in similar content.

Over 35\% of the accounts that have been exposed to our content choose to view it. Figure \ref{fig:accountsreached} (top) illustrates both the impressions and accounts reached. On Instagram, 'impressions' refer to the number of times our content has been displayed to users, meaning the total number of opportunities for an account to see a video, post, or account (e.g., through the search feed). 'Accounts reached' represents the number of accounts that, after being prompted by Instagram, actually decided to view the content. However, this does not guarantee that they watched the videos in their entirety or interacted with them.

\begin{figure}[bt]
\includegraphics[width=0.45\textwidth]{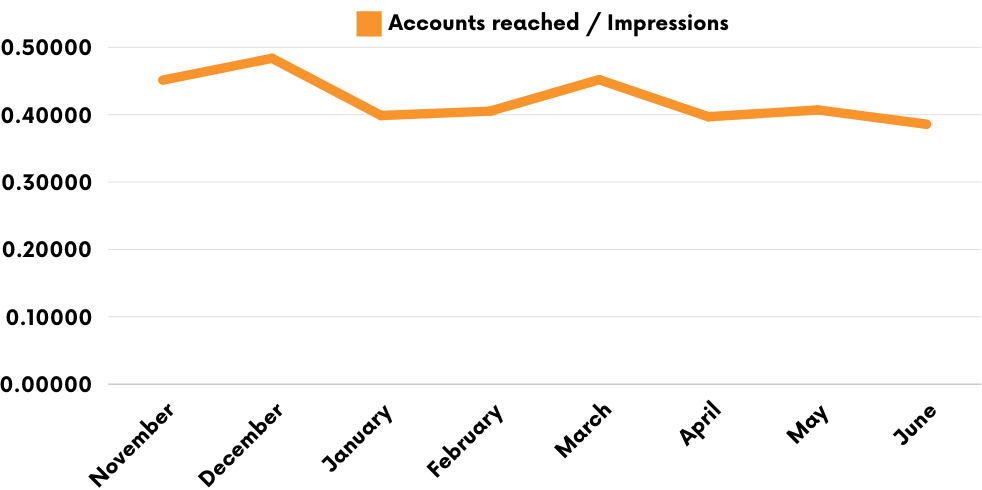}
\includegraphics[width=0.45\textwidth]{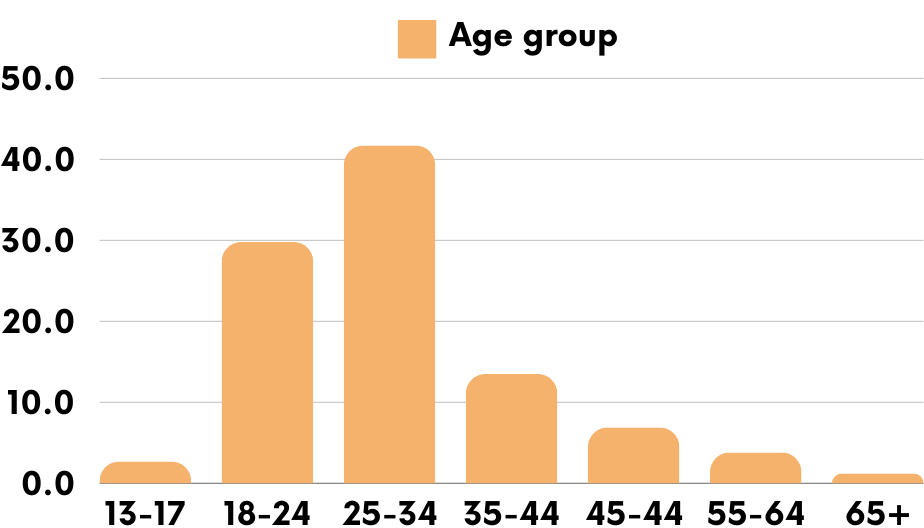}
\centering
\caption{\label{fig:ratioimpressions} (Left): Ratio between the total accounts reached and the impressions from November 2023 to June 2024. (Right): Percentage of followers corresponding to different age groups in our community. Data collected in June 2024.  }
\label{fig:impressions-ages-com}
\end{figure}

In Figure \ref{fig:accountsreached}, we present the ratio between the accounts reached and the impressions. The ratio consistently remains above 35\%. The slow decrease in recent months is attributable to the substantial increase in impression. As Instagram recognizes the value of our content for specific users, the platform displays it more frequently and to a larger audience. This is common in social media, when the absolute number of accounts choosing to view the content rises (Figure \ref{fig:impressions-ages-com}), the percentage of users who engage with it tends to decrease.

%\subsection{The Community}

When analyzing follower numbers, it is clear the target audience is reached successfully. The key indicator is the followers' age distribution. As shown in Fig.~\ref{fig:impressions-ages-com}, more than 40\% of followers are between 25 and 34 years old, and over 70\% fall within the 18 to 34 age range.
The audience reach is predominantly concentrated in Germany, accounting for about 30\% of followers. However, the account has significant reach in other countries. Additionally, our follower demographics show an approximate 1:3 ratio of women to men (Figure \ref{fig:community}). In the central/northern European STEM context, these numbers are promising\ \cite{shefigures}. 

\iffalse
\begin{figure}[bt]
\begin{center}
 \setkeys{Gin}{}
    \adjustboxset{scale=0.53, valign=c}
 \begin{tabular}{ccc}
\adjustimage{}{graphs/com-countries.png} &  
 & \includegraphics[width=0.39\textwidth]{graphs/com-gender.png}\\ 
\end{tabular}
\end{center}
 \caption[Demography in our community.]{Demographic statistics present in our community including the countries where the accounts are located and user's gender. Data collected on June 2024.}
 \label{fig:community}
\end{figure}
\fi

\begin{figure}[bt]
\includegraphics[width=0.3\textwidth]{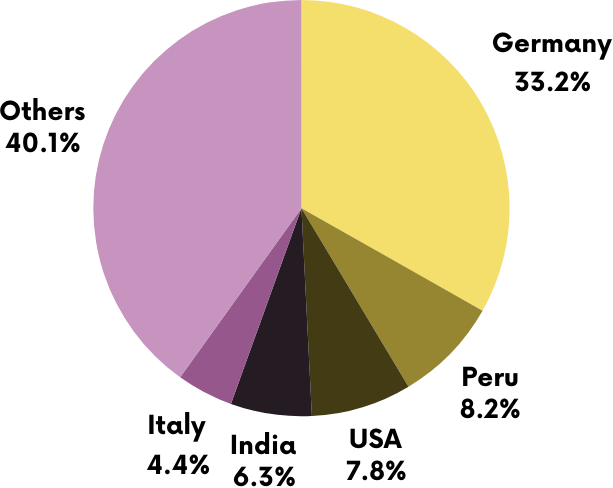}\hspace{0.3cm}
\raisebox{0.5cm}{\includegraphics[width=0.3\textwidth]{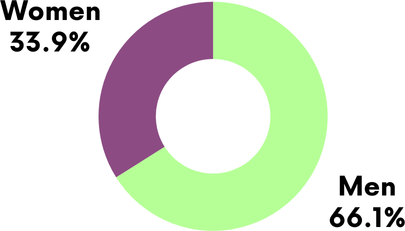}}
\centering
\caption{\label{fig:ratioimpressions} Demographic statistics present in our community including the countries where the accounts are located and user's gender. Data collected on June 2024.}
\label{fig:community}
\end{figure}

\section{Conclusion}

The crowd-sourced particle physics stories from CMSatDESY feature a diverse team of scientists working collaboratively to manage an Instagram account dedicated to sharing our personal stories and experiences. The initiative serves four primary objectives: outreach, science communication, training for local individuals, and recruitment efforts. The account has shown consistent growth and is successfully reaching the target audiences. Its concept and organisation ensure that it remains sustainable now and in the future, effectively combining diverse expertise with a strategic approach to engaging and educating the public.

\end{document}